\documentclass[10pt, preprintnumbers,amsmath,amssymb]{revtex4}

\usepackage{latexsym}
\usepackage{amssymb}
\usepackage{epsfig,amsmath,graphics}
\usepackage{epstopdf}
\usepackage{comment}
\usepackage{float}
\usepackage{ulem}
\usepackage{cancel}

\newcommand{\BPV}{$\text{\cancel{B}RPV\,}$}

\begin{document}

\title{Baryonecrosis: Displaced Vertices from R-Parity Violation}

\author{Kurt Barry}
\affiliation{Stanford Institute for Theoretical Physics, Department of Physics, Stanford University, Stanford, CA 94305}

\author{Peter W. Graham}
\affiliation{Stanford Institute for Theoretical Physics, Department of Physics, Stanford University, Stanford, CA 94305}

\author{Surjeet Rajendran}
\affiliation{Stanford Institute for Theoretical Physics, Department of Physics, Stanford University, Stanford, CA 94305}

\begin{abstract}
The LHC has placed stringent limits on superpartner masses, in conflict with naturalness.  R-parity violation is one of the few scenarios that allows reduction of these limits and is thus worth significant exploration at the LHC.  We demonstrate that if the R-parity violating operator $UDD$ is used, we generically expect all SUSY events at the LHC to have displaced vertices.  If a squark is the LSP, it will have a short displaced vertex.  If any other particle is the LSP, the displaced vertex is naturally expected to be quite long, possibly even outside the detectors.  These scenarios are already constrained by existing searches for missing energy. This arises because this operator efficiently washes-out the baryon asymmetry in the early universe, unless the squarks are heavy and the coupling is small.  Avoiding displaced vertices is possible, but requires baryogenesis below the weak scale.  Thus, for example, the use of sphalerons in baryogenesis does not avoid the requirement of displaced vertices.  This motivates searching for hadronic displaced vertices at the LHC with decay lengths anywhere from tens of microns to meters.
\end{abstract}

\maketitle

\tableofcontents

\section{Introduction}
\label{Sec:Introduction}
The most prominent searches for supersymmetry (SUSY) at the Large Hadron Collider are based on the expectation that every SUSY particle produced in a collision rapidly cascades down to the lightest supersymmetric particle (LSP). The LSP is assumed to be stable and neutral and is expected to leave the collider without further interaction,  giving rise to missing energy signals. The LHC has severely challenged this scenario, drastically constraining the parameter space where these models can have light super-partners that solve the hierarchy problem \cite{Craig:2013cxa, Buchmueller:2013rsa}. These bounds can be significantly relaxed if the LSP decays inside the collider to jets and charged leptons, as such decays can significantly decrease or even completely remove missing energy from the event \cite{KaplanUDD, Carpenter:2008sy, Allanach:2012vj}. 

The stability of the LSP is not a requirement of the SUSY algebra, but is rather a consequence of the imposition of R-parity. This parity was introduced to protect the stability of the proton, since its absence allows for the existence of low-energy baryon and lepton number violating operators that lead to proton decay. But, since the proton can decay only when both baryon and lepton number are violated, it is possible to break R-parity while conserving one of these numbers, leading to the decay of the LSP while preserving the stability of the proton \cite{Hall:1983id} and thus relaxing the constraints on SUSY. Of these scenarios, R-parity violation that breaks lepton number is also significantly constrained since it gives rise to lepton-rich signals that can usually be searched for relatively easily in a hadron collider, although specific realizations of this scenario could relax these bounds \cite{Graham:2012th}. But, R-parity violation that breaks baryon number ($\text{\cancel{B}RPV}$) predominantly produces events  with jets in the final state. These well-motivated scenarios are experimentally more challenging to detect since they need to be distinguished from the large QCD background in the LHC environment, making this one of the most difficult channels in which to search for SUSY at the LHC \cite{Krnjaic:2013eta, Brust:2011tb, Brust:2012uf,  Bai:2013xla, Duggan:2013yna, Berger:2013sir, Franceschini:2012za, Asano:2012gj, Evans:2012bf, Baer:2013faa}. 

In this paper, we point out that \BPV operators will generically lead to displaced vertices at the LHC, with displacement lengths that are order $\sim 100 \, \mu$m to m. This expectation arises from the fact that for the \BPV operators to alleviate the constraints on SUSY at the LHC, the LSP has to decay within the collider, with the decay length inversely related to the strength of the \BPV operator. But, a large \BPV operator can erase the baryon number of the universe, eliminating our existence. We show that the resolution of this tension generically leads to collider-scale displaced vertices. While it has long been recognized that the size of \BPV operators is constrained by the preservation of baryon number \cite{Dreiner:1992vm, Barbier:2004ez, An:2013axa, Campbell:1990fa, Bouquet:1986mq}, the collider-scale displaced vertices at the LHC implied by this bound have not yet been appreciated. While new techniques might be necessary to specifically identify these vertices \cite{Graham:2012th}, they provide an interesting handle for distinguishing these SUSY events from the QCD background.  

We begin in Section \ref{Sec:Synopsis} by illustrating the basic argument for the connection between cosmological baryon preservation and displaced vertices at the LHC. This argument depends upon the temperature at which 
the baryon asymmetry is introduced into the universe. The baryon washout processes induced by the \BPV operators are ineffective at low temperatures. We compute the highest temperature which preserves baryon number in 
Section \ref{Sec:Methods}, starting with a qualitative discussion in Section \ref{GreatExpectations}. While displaced vertices can be avoided if baryogenesis occurs at temperatures lower than the results in Section 
\ref{Sec:Results}, these temperatures are low enough to rule out most models of baryogenesis. This includes electroweak baryogenesis, which is known to be excluded in the MSSM based on the observed Higgs 
properties \cite{Curtin:2012aa,Cohen:2012zza}. Though baryogenesis can be performed at these very low temperatures \cite{Dimopoulos:1987rk, Affleck:1984fy, Cui:2012jh, Davidson:1996hs}, given the generic ease with which it can be 
accommodated at high temperatures, there exists an excellent case to search for \BPV SUSY in such collider-scale displaced vertices. If naturalness is sacrificed, our bounds provide motivation for models of baryogenesis 
involving the \BPV coupling in which squarks are very heavy (e.g.\ \(> 100 \text{ GeV}\)); such scenarios may still exhibit displaced vertices, as in \cite{Cui:2013bta}.

\section{Synopsis}
\label{Sec:Synopsis}
In the presence of \BPV operators, baryon number is no longer conserved and thermal processes will restore baryon number to the thermal equilibrium value of zero. These baryon destruction processes will remain in equilibrium until a temperature below which the Hubble scale exceeds the rate of baryon destruction. If baryon number is introduced below this temperature, the baryon destruction processes are not in equilibrium and the introduced asymmetry will be preserved. 

The \BPV operators of the MSSM which are of the form $\lambda''_{ijk}U_{i}D_{j}D_{k}$ violate baryon number through the processes listed in Figure \ref{AllProcs}, all of which require either the presence or the creation of squarks. These are the most efficient channels of baryon destruction. When the squarks are off-shell, they can be integrated out yielding baryon number violating higher dimension operators in the Standard Model. But, the operators of lowest dimension that violate just baryon number in the standard model are dimension 9 operators of the form $UDDUDD$ (leading to, for example, neutron-anti neutron oscillations). These dimension 9 operators rapidly decouple and are not efficient. Since efficient baryon destruction through the \BPV operators requires the presence of on-shell squark degrees of freedom, they will be ineffective below temperatures $T \lesssim\frac{m_{\tilde{q}}}{20}$ of the squark mass $m_{\tilde{q}}$. A natural SUSY spectrum requires $m_{\tilde{q}} \sim \mathcal{O}\left(100 - 1000\right) \text{ GeV}$, implying that these baryon number violating processes will be effective at least until temperatures $T \sim \mathcal{O}\left(100 \text{ GeV}\right)$.

The complete destruction of baryon number through these \BPV operators also requires an efficient mechanism to transfer baryon number between quarks and squarks. This is necessary since the \BPV operators preserve the baryon number in quarks minus twice the baryon number in squarks. This number is, however, broken by gaugino interactions, such as the gluino exchange diagram in Figure \ref{ExchangeProc}. For a natural SUSY spectrum, the gaugino masses must also be $\sim \mathcal{O}\left(\text{TeV}\right)$, making these processes much more rapid than the baryon destruction caused by the \BPV operators because they are not suppressed by the smallness of the $\lambda''$ couplings.

The Hubble scale at \(T \sim \mathcal{O}\left(100 \text{ GeV}\right)\) is $\sim \mathcal{O}\left(10^{9}\text{ s}^{-1}\right) $. The typical rate for the baryon destruction processes in Figure \ref{AllProcs} scales as $\sim \left|\lambda_{ijk}^{''}\right|^{2} m_{\tilde{q}}$. Preservation of baryon number thus requires this rate to be smaller than the Hubble scale at the decoupling temperature {\it i.e.} $ \left|\lambda_{ijk}^{''}\right|^{2} m_{\tilde{q}} \lesssim \mathcal{O}\left(10^{9}\text{ s}^{-1}\right)$. 

At the LHC, in the presence of \BPV operators, the LSP will decay at a rate that is no slower than $\sim  \left|\lambda_{ijk}^{''}\right|^{2} m_{\tilde{q}}$, since the phase space available for the decay cannot be larger than a typical squark mass $m_{\tilde{q}}$ and the decay has to involve the \BPV couplings $\lambda''_{ijk}$. This implies that the decay rate of the LSP has to be slower than $\mathcal{O}\left(10^{9}\text{ s}^{-1}\right)$, which leads to displaced vertices at the LHC with decay lengths $\sim 100 \, \mu$m to $1$ m depending upon the exact decay topology. 

Thus we see that the preservation of baryon number naturally leads to collider-scale displaced vertices in \BPV models of low-energy SUSY, unless baryon number is introduced at temperatures well below $\sim 50$ GeV (see section \ref{Sec:Results}). 

\section{Analysis}
\label{Sec:Methods}

\subsection{Expected Qualitative Behavior}
\label{GreatExpectations}

The qualitative behavior of the boundary between the baryon survival and destruction regions in the squark mass versus \(\log{|\lambda''|}\) plane (see figure \ref{DeathRegionMassVsLambdaSquark}) may be deduced from a physical argument. Let \(T\) be the temperature at which a baryon asymmetry is 
introduced. The threshold condition for survival of this asymmetry, very roughly speaking, is when baryon-destroying processes freeze out at the temperature of introduction. The freeze-out condition is that the rate of these processes is of order one compared to Hubble. The rate is generally expressible as a 
number density times a cross section:
\begin{equation}
  n \langle \sigma v \rangle \sim H.
\end{equation}
For starting temperatures well below the squark mass \(m_{\tilde{q}}\), \(n \langle \sigma v \rangle\) will generically be exponentially damped by a factor of \(e^{-m_{\tilde{q}}/T}\); substituting this parametric dependence, along with that for Hubble and a few factors to keep the units consistent, one obtains:
\begin{eqnarray}
 \nonumber
 (m_{\tilde{q}}T)^{(3/2)} e^{-m_{\tilde{q}}/T} \frac{|\lambda''|^2}{m_{\tilde{q}}^2} & \sim & \frac{T^2}{M_{Pl}} \\
 -\frac{m_{\tilde{q}}}{T} + 2 \ln{|\lambda''|} & \sim & \ln{\frac{\sqrt{m_{\tilde{q}}T}}{M_{Pl}}}.
\end{eqnarray}
The right-hand side is approximately a constant so one obtains:
\begin{equation}
  m_{\tilde{q}} \sim \frac{2T}{\log_{10}{(e)}}\log_{10}{|\lambda''|} - T\ln{\frac{\sqrt{m_{\tilde{q}}T}}{M_{Pl}}}.
\end{equation}
This is only a rough approximation, since the exponential damping of squark number density (or the thermal average in the case of processes without initial state squarks) does not fully come into effect until temperatures far below the squark mass (e.g.\ \(m_{\tilde{q}}/20\)).  However, the qualitative behavior (a 
straight line) is confirmed, and the formula for the slope presented above becomes increasingly accurate for smaller starting temperatures. For example, at 100 GeV the analytic result for the slope differs by 21\% from the true value; for a 10 GeV starting temperature, the error is only 4\% (see figure 
\ref{TempContours}). The predicted value for the y-intercept is also reasonable. Choosing a mass of 700 GeV yields an estimate of \(\sim\)3800 GeV for a 100 GeV initial temperature; the actual value (which can be estimated from figure \ref{DeathRegionMassVsLambdaSquark}), is \(\sim\)4400 GeV.

For a calculation beginning and ending at temperatures well above \(m\), the squark mass is irrelevant and one obtains a vertical line at a cutoff value of \(\lambda'' \sim \sqrt{T/M_{Pl}}\), which can be derived by inserting the 
parametric behavior \(n\langle \sigma v \rangle \sim T|\lambda''|^2\) into the chain of reasoning above. If such a calculation is extended down to temperatures below the squark mass until baryon number is destroyed (as in this paper), then for lower squark masses the 
baryon destruction will be more effective, resulting in a positive slope for the boundary curve as a function of \(\log{|\lambda''|}\). This expected behavior is also observed. The predicted cutoff for our chosen high-scale initial temperature of 12 TeV is 
\(|\lambda''| \sim \sqrt{12\cdot10^3/10^{19}} \sim 10^{-7.5}\); the observed (positively sloped) cutoff occurs at roughly \(10^{-7.7}\) for \(m_{\tilde{q}} = 200\text{ GeV}\) and \(10^{-7.5}\) for \(m_{\tilde{q}} = 1200\text{ GeV}\) (see figure \ref{DeathRegionMassVsLambdaSquark}).

\subsection{Processes for Baryon Washout}
\begin{figure}
  \centering
  \includegraphics[scale=0.8]{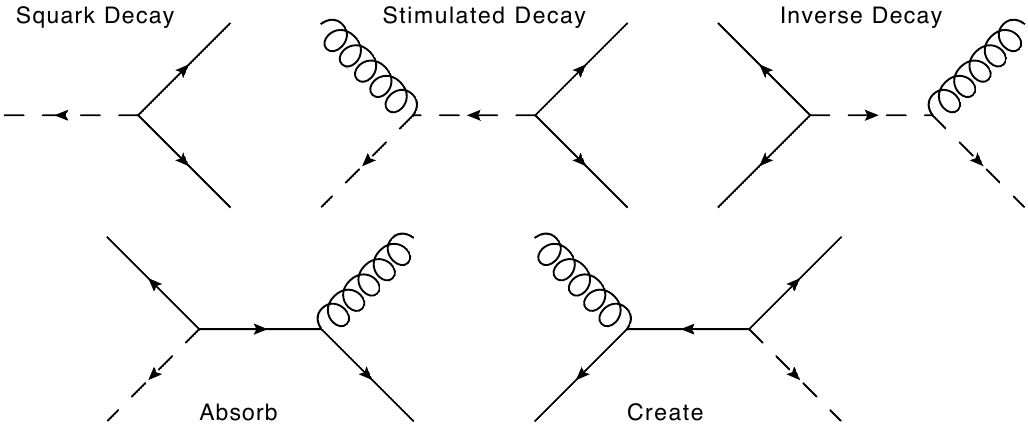}
  \caption{The primary processes contributing to baryon number destruction. They have been named according to what happens to the squark in each process. Solid lines 
  are quarks, dashed lines are squarks, and the curly lines may be taken either as gluons or as other appropriate gauge bosons.}
  \label{AllProcs}
\end{figure}

\begin{figure}
  \centering
  \includegraphics[scale=0.8]{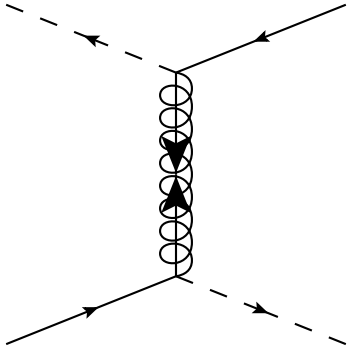}
  \caption{The exchange process that transfers baryon number between the squark and quark sectors.}
  \label{ExchangeProc}
\end{figure}

There are five \BPV processes that are of primary importance in baryon washout, shown in Figure \ref{AllProcs}. At temperatures well above the squark mass, the two-to-two scattering 
processes are dominant, due to enhancement from greater particle number densities and time dilation suppression of the decay process. At temperatures below the squark mass, the decay 
process begins to dominate, though the two-to-two processes cannot be neglected. Note that the Stimulated Decay and Inverse Decay processes are related by time reversal symmetry, as are the 
Absorb and Create processes. For each two-to-two process, there are two additional Feynman diagrams which are not displayed. There is one other important process (mentioned in Section \ref{Sec:Synopsis}), 
involving a virtual gluino, that allows the exchange of baryon number between the quark and squark sectors (see figure \ref{ExchangeProc}). Without this process, the difference in baryon 
asymmetry between quarks and twice that of squarks will be conserved.  The gluino does need to be very light to make squark-quark interconversion rapid enough to be effectively infinite. For a 1500 GeV 
gluino, exchanges are already up to seven orders of magnitude more rapid than the typical rate of baryon destroying processes (for relevant values of the \BPV coupling--those close to 
the transition between baryon survival and destruction). As discussed previously, a gluino mass not too different from the squark mass is motivated by naturalness.

Matrix elements were computed using the Weyl spinor formalism described in \cite{Martin}, and thermal averaging was based on the methods of \cite{GonzoAndJellybean}. Exact formulae for the 
matrix elements for the Stimulated Decay and the Absorb processes are given in the appendix.

\subsection{Evolution Equation and Approximations}

The evolution of baryon number is tracked through the Boltzmann equation. In order to make these equations tractable, two major simplifying assumptions were made. First, all squarks were assumed to be equally massive. Second, it was assumed that all of the \BPV couplings \(\lambda''_{ijk} U_i D_j D_k\) were identical for all values of the family indices \(ijk\). This is reasonable since baryon survival is only logarithmically sensitive to these parameters, so only their order-of-magnitude truly matters. Existing bounds on \BPV couplings also do not greatly constrain this assumption. The strictest are from the nucleon decay, but these all involve other R-parity violating interactions besides \(UDD\), and thus these can be made arbitrarily small to compensate for \(UDD\) largeness. The only significant constraint is neutron-antineutron oscillation, which limits \(\lambda''_{121} \lesssim 10^{-6}\) for a SUSY-breaking mass of 300 GeV, but does not constrain any other couplings \cite{Barbier:2004ez}.

\begin{figure}
  \centering
  \includegraphics[scale=0.8]{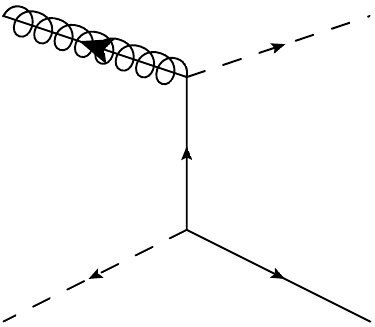}
  \caption{On shell gluino processes like this one also contribute to baryon number destruction at high temperatures, ensuring that our result is conservative.}
  \label{OnShellGluino}
\end{figure}

The following additional approximations were also used.  Those that may have a significant effect or were critical in simplifying the calculations are listed and discussed here. The more minor ones are can be found in the appendix.
\begin{itemize}
\item
  Above the weak scale (roughly 100 GeV), electroweak vacuum transitions involving sphalerons violate B+L, creating additional complexity in the baryon number dynamic.  These may generate a baryon asymmetry for the universe if the sphalerons are active after the baryon violating RPV interactions go out of equilibrium (see for example \cite{Dreiner:1992vm}).  Such a mechanism can happen for example in leptogenesis \cite{Fukugita:1986hr}.  However this requires the sphalerons to be active at temperatures below where the \BPV interactions have shut off.  If the \BPV interactions are in equilibrium at temperatures below the sphalerons, the \BPV interactions will force the baryon asymmetry to zero.  Sphalerons are known to be inefficient below a temperature around the weak scale, between 140 and 175 GeV \cite{D'Onofrio:2012jk}.  Thus for all of our lower scale results the effect of sphalerons is negligible.  For our higher scale results, when the initial temperature is taken to be well above this scale, the sphalerons could possibly regenerate the baryon asymmetry (e.g.~for the 12 TeV regions of Figures \ref{DeathRegionMassVsLambdaSquark}, \ref{DeathRegionMassVsLambdaNeutralino}, and \ref{DeathRegionMassVsLambdaNeutralino2}).  Sphalerons will only be able to regenerate the baryon asymmetry given the right initial conditions, e.g.~a lepton flavor asymmetry that persists below where the \BPV interactions shut off.  We neglect the sphalerons in our calculation.  Our main results still hold since even just using the effect of \BPV interactions below 140 GeV where the sphalerons are negligible leads to displaced vertices for SUSY events.

\item
  All quarks have been treated as massless. This is correct at temperatures above the electroweak phase transition, and is only an issue for the top quark at lower temperatures (the 
  low-temperature cutoff is 1 GeV). If one does not make this assumption, it is necessary to track the top number separately, which creates  additional complexity.

\item
  Quantities depending on chemical potential were expanded to first order under the assumption \(\mu/T < 1\). If \(n\) is the number density of some species, and \(\bar{n}\) the number 
  density of its antiparticle, this expansion allows the crucial approximation \(n+\bar{n} \approx 2n^{EQ}\), where \(n^{EQ}\) is the equilibrium number density of that species. Without 
  this substitution, the Boltzmann equations cannot be written purely in terms of baryon asymmetries, and are thus much more complicated. The value of the initial baryon asymmetry was 
  chosen to limit inaccuracies from this choice to less than 1\%.
  
 \item Rapid exchange processes mediated by \(SU(2)_L\) interactions combined with Higgs exchange result in identical baryon asymmetries across different quark fields implying that the net baryon asymmetry in the quark sector can be tracked with a single function.  

 \item
  Diagrams involving initial and final state gluinos were neglected (see figure \ref{OnShellGluino}). Since we assumed gluinos to be significantly heavier than squarks (at least 1500 GeV), this is negligible at low 
  temperatures. However, the gluino cannot be made arbitrarily heavy if one desires naturalness. Thus, at temperatures above a few hundred GeV, processes involving on-shell gluinos in the initial and final states
  will enhance baryon destruction. A rough way to (over)estimate the effect of gluinos is to simply double the rates of the two-to-two processes (since supersymmetry allows the replacement 
  of gluons with gluinos and quarks with squarks in all the diagrams). This is equivalent to a shift in \(\log{\lambda}''\) of \(\sqrt{2}\) at high temperatures, an effect that is 
  noticeable but not overly large.

\end{itemize}

\begin{table}
\begin{tabular}{|c|c|}
\hline
  \textbf{Variable}                    & \textbf{Description} \\ \hline
  \(m_{\tilde{q}}\)                                & squark mass \\ \hline
  \(z\)                                & $m_{\tilde{q}}$/T; evolution variable \\ \hline
  \(n^{EQ}_{\tilde{q}}\)               & squark equilibrium number density \\ \hline
  \(n^{EQ}_{R}\)                       & equilibrium number density of a massless species \\ \hline
  \(q\)                                & asymmetry of one quark degree of freedom divided by \(n^{EQ}_{R}\) \\ \hline
  \(\langle \Gamma \rangle\)           & thermally averaged squark decay rate \\ \hline
  \(\langle \sigma_{SD}|v| \rangle\)   & thermally averaged cross section times velocity for the Stimulated Decay process \\ \hline
  \(\langle \sigma_{ID}|v| \rangle\)   & thermally averaged cross section times velocity for the Inverse Decay process \\ \hline
  \(\langle \sigma_{C}|v| \rangle\)    & thermally averaged cross section times velocity for the Create process \\ \hline
  \(\langle \sigma_{A}|v| \rangle\)    & thermally averaged cross section times velocity for the Absorb process \\ \hline
  $H(m_{\tilde{q}})$ & Hubble parameter when the temperature is equal to the squark mass $m_{\tilde{q}}$ \\ \hline

\end{tabular}
\caption{Definitions of the variables used in the evolution equation \eqref{eqn: boltzman eqn}.}
\label{vartab}
\end{table}

As a result of these approximations, the Boltzmann equation governing the evolution of the baryon asymmetry is 
\begin{eqnarray}
  \nonumber
  \frac{d}{dz}\left( \left(1+\frac{n^{EQ}_{\tilde{q}}}{n^{EQ}_R}\right)q \right) & = & -z \left[ \frac{27}{4}\frac{\langle \Gamma \rangle}{H(m_{\tilde{q}})}\frac{n^{EQ}_{\tilde{q}}}{n^{EQ}_R}q \right. \\ \nonumber
   &&  + 108\frac{\langle \sigma_{SD}|v| \rangle}{H(m_{\tilde{q}})}n^{EQ}_{\tilde{q}}q + \frac{81}{2}\frac{\langle \sigma_{ID}|v| \rangle}{H(m_{\tilde{q}})}n^{EQ}_{R}q \\ 
   &&  \left. + 192\frac{\langle \sigma_{C}|v| \rangle}{H(m_{\tilde{q}})}n^{EQ}_{R}q + 72\frac{\langle \sigma_{A}|v| \rangle}{H(m_{\tilde{q}})}n^{EQ}_{\tilde{q}}q \right]
   \label{eqn: boltzman eqn}
\end{eqnarray}
where  \(z \equiv m_{\tilde{q}}/T\), where \(m_{\tilde{q}}\) is the common squark mass  and an explanation of each variable that appears is given in Table \ref{vartab}. Note that only terms arising from QCD processes are shown for the two-to-two terms. An important subtlety is accounting for the family factors arising from the \(UDD\) coupling indices (the \(ijk\) on \(\lambda''_{ijk}\)). These are different between squarks and quarks, and also between up-type and down-type species, and even the specific processes.

The value of baryon number density per unit relativistic number density at a temperature of 1 GeV needs to be roughly \(10^{-8}\) to yield the correct value for today. This quantity at any 
stage of the calculation may be obtained as follows:
\begin{equation}
  B = 12(q + (n^{EQ}_{\tilde{q}}/n^{EQ}_R)q)
\end{equation}
The first term is for quarks, the second is for squarks. To obtain the prefactor, one must account for family, color, baryon charge (1/3 for both squarks and quarks), and the set of \(U\), \(D\), \(U^c\), \(D^c\). Since baryon number evolution is monotonic, baryons are considered ``destroyed'' once this quantity drops below the value it holds in our universe.

\section{Results}
\label{Sec:Results}

\begin{figure}
  \centering
  \includegraphics[scale=0.5]{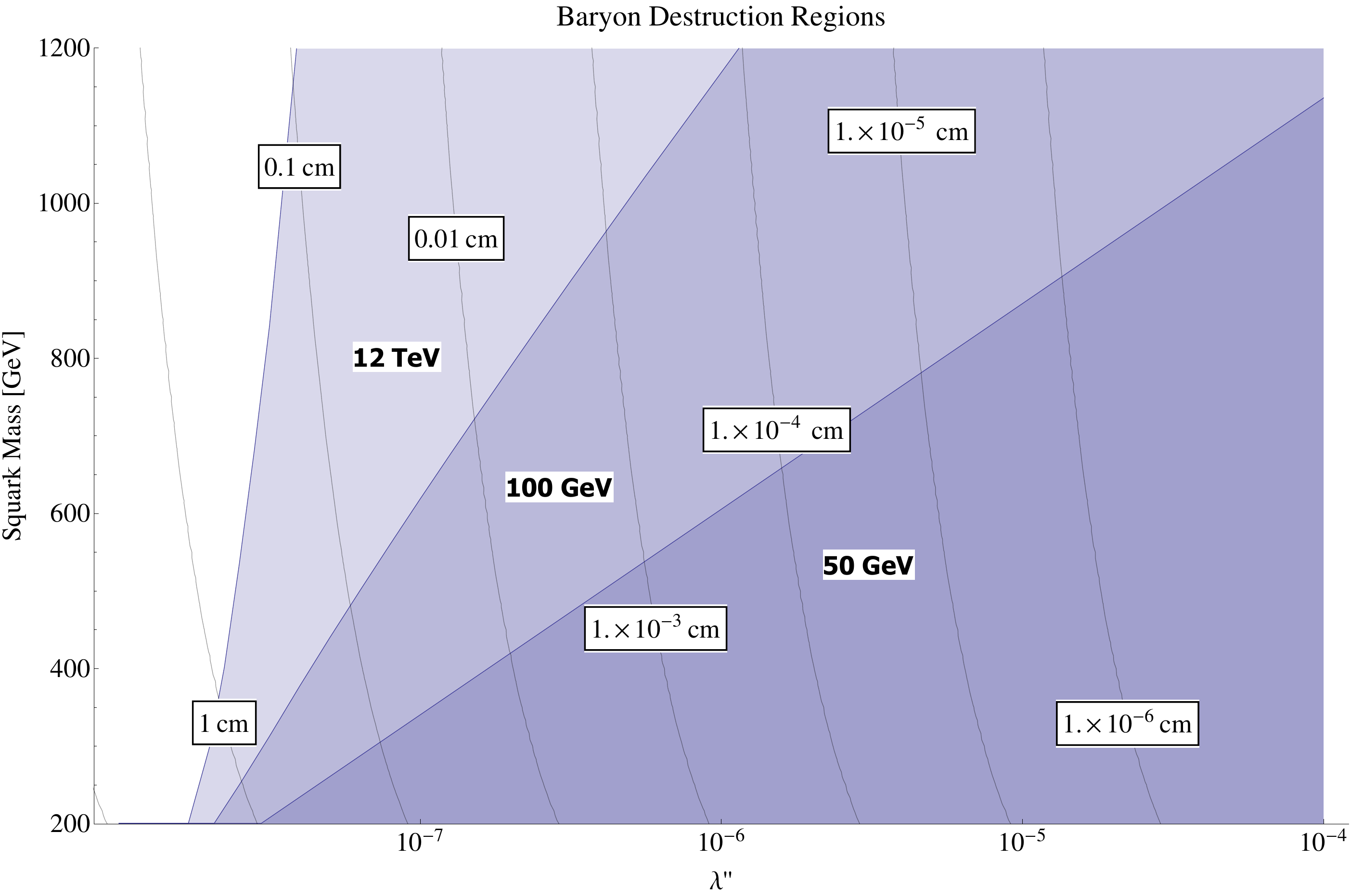}
  \caption{This figure shows the regions of parameter space (shaded in blue) in which the baryon asymmetry is washed out by the \BPV processes.  Each region is labeled by the temperature at which an order-one baryon asymmetry is assumed to be initially generated. The horizontal axis is the \BPV \(UDD\) coupling.  The overlayed contours give the decay length of the LSP in centimeters. As an example, if the squark mass is 400 GeV and $\lambda'' = 10^{-7}$, the universal baryon asymmetry must be generated at 
  a temperature of less than 100 GeV, but may be generated above 50 GeV, and the resulting squark decay length will be between 0.01 and 0.1 centimeters. The qualitative features of these regions are predicted by the arguments of Section \ref{GreatExpectations}.}
  \label{DeathRegionMassVsLambdaSquark}
\end{figure}

\begin{figure}
  \centering
  \includegraphics[scale=0.5]{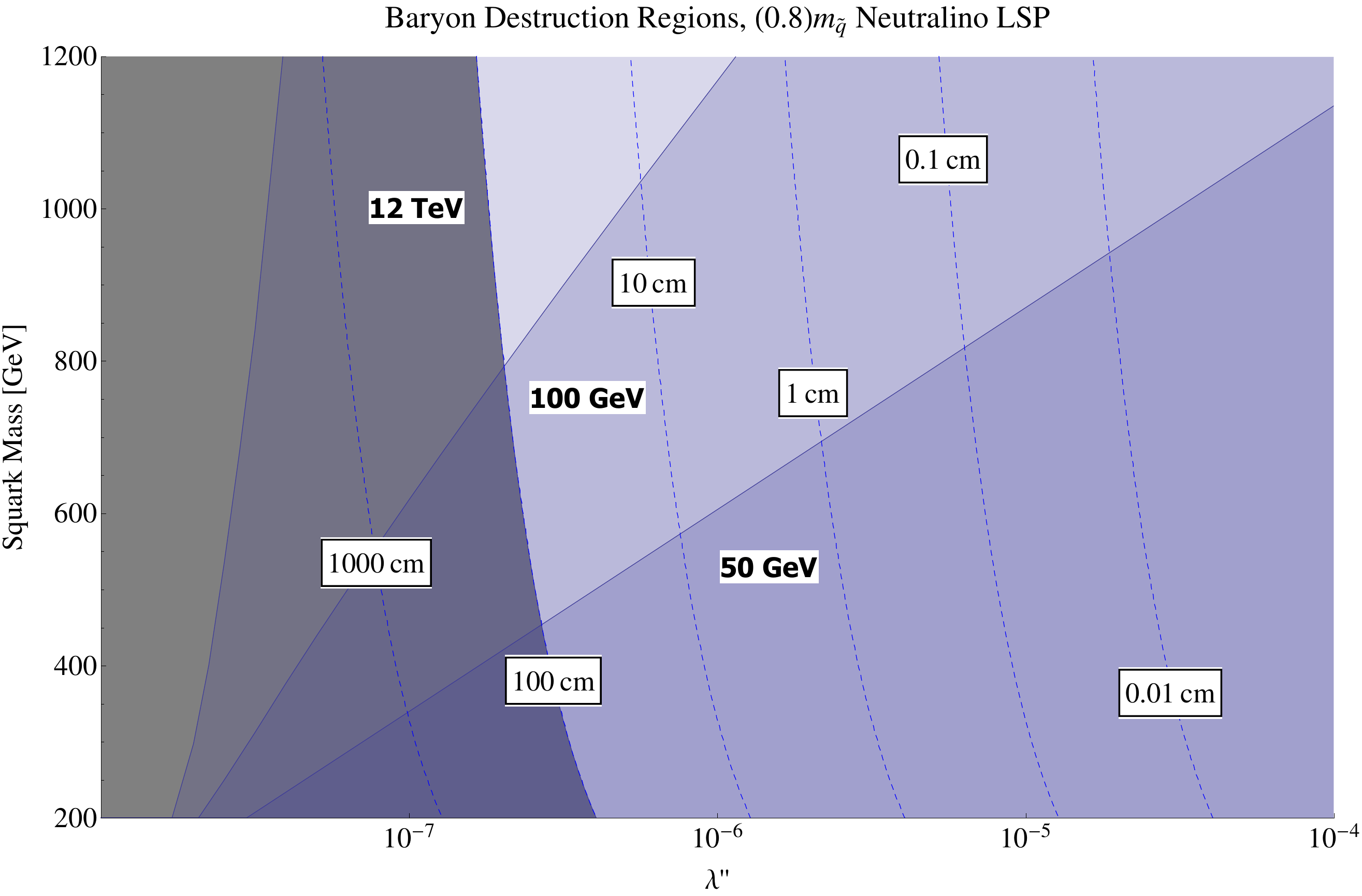}
  \caption{This figure shows the same excluded regions as Figure \ref{DeathRegionMassVsLambdaSquark}. Here, however, a neutralino LSP with mass 0.8 $m_{\tilde{q}}$ is assumed, and the contours correspond  to the expected neutralino decay lengths, in centimeters. The allowed neutralino decay lengths are generically much longer than those for a squark LSP.  Neutralino decay lengths longer than $\sim 1$ m are likely ruled out by missing energy searches at the LHC, as shown in the (gray) shaded region.}
  \label{DeathRegionMassVsLambdaNeutralino}
\end{figure}

\begin{figure}
  \centering
  \includegraphics[scale=0.5]{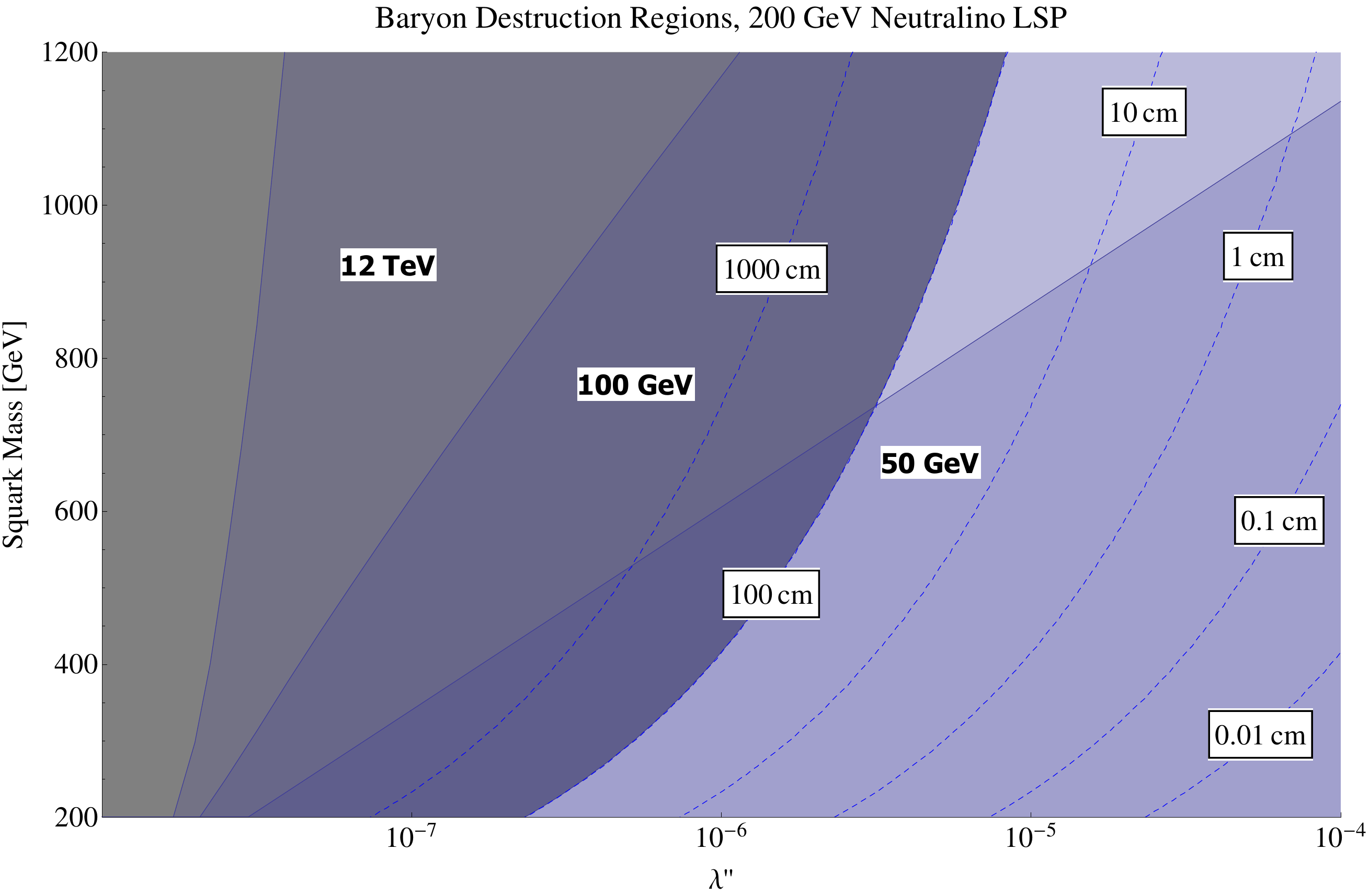}
  \caption{Similar to Figure \ref{DeathRegionMassVsLambdaNeutralino}, this figure shows the same excluded regions as Figure \ref{DeathRegionMassVsLambdaSquark}. Here, however, a neutralino LSP with mass = 200 GeV is assumed, and the contours correspond  to the expected neutralino decay lengths, in centimeters. The allowed neutralino decay lengths are generically much longer than those for a squark LSP.  Neutralino decay lengths longer than $\sim 1$ m are likely ruled out by missing energy searches at the LHC, as shown in the (gray) shaded region.}
  \label{DeathRegionMassVsLambdaNeutralino2}
\end{figure}

Figures \ref{DeathRegionMassVsLambdaSquark}, \ref{DeathRegionMassVsLambdaNeutralino}, and \ref{DeathRegionMassVsLambdaNeutralino2} display our main results.  They show the regions of the squark mass versus RPV coupling $\lambda''$ plane in which the baryon number violation is strong enough to completely wipe out the baryon asymmetry of the universe (even with an order-one initial asymmetry).  The shaded regions are excluded, and are labeled by the temperature at which the baryon asymmetry is assumed to be generated at order one.  Thus in order to have our observed baryon asymmetry inside one of these regions, the baryon asymmetry must be generated below the labeled temperature for that region.
We assume that there is an order-one initial baryon asymmetry.  Such a large asymmetry is difficult to get from most baryogenesis models so this assumption is conservative.  A smaller initial asymmetry would lead to even more stringent constraints than shown.  However our limits are quite insensitive (only logarithmically) to the initial baryon asymmetry because once the baryon number violation is in thermal equilibrium the initial asymmetry is washed out very efficiently.
The qualitative arguments of Section \ref{GreatExpectations} explain why straight lines occur for 
the low-temperature regions and why the high temperature region is approximately a simple cutoff at \(\log|\lambda''| \sim -7.5\).

The overlayed contours in Figure \ref{DeathRegionMassVsLambdaSquark} show the decay length for a squark LSP (in units of centimeters) as a function of \(\log{|\lambda''|}\) and squark 
mass.  Figure \ref{DeathRegionMassVsLambdaNeutralino} shows the same exclusion regions as figure \ref{DeathRegionMassVsLambdaSquark}, but is overlayed with (approximate) decay length contours assuming a neutralino LSP whose mass is fixed to be $0.8$ of the squark mass $= 0.8 \, m_{\tilde{q}}$.  We take this value because it is the closest to the squark mass that seems at all reasonable from a natural point of view.  This makes the neutralino decay length as short as possible.  If the neutralino LSP is lighter the decay length will be longer.  Thus the neutralino LSP decay length contours are conservative, and the decay length is likely to be even longer in this scenario.  Figure \ref{DeathRegionMassVsLambdaNeutralino2} shows the same plot but with a neutralino with mass fixed to be 200 GeV.  In Figures \ref{DeathRegionMassVsLambdaNeutralino} and \ref{DeathRegionMassVsLambdaNeutralino2}, we have also shaded the region where the neutralino decay length is $\gtrapprox 1$ m. In this case, the event will appear as missing energy at the LHC since the decaying neutralino will not leave a track in the tracker and will be reconstructed as missing energy. This part of parameter space is already constrained by existing LHC searches, unless baryogenesis was performed at a temperature lower than the weak scale.  Note that other types of LSP's would also have decay lengths as least as long as the neutralino, if not longer.

From these figures it is clear that if the baryon asymmetry is generated anywhere around or above the weak scale, then we will generically expect the LSP to decay with a displaced vertex.  This is because the RPV coupling must be quite small in order to avoid destroying the baryon asymmetry.  If the LSP is a squark the decay length is generically expected to be a cm or longer.  If the LSP is a neutralino the decay length is expected to be much longer, on the order of several meters.  In fact, generically a neutralino LSP may have such long decay lengths that a fraction of the events will appear as missing energy at the LHC.  Such events would have shown up in MET searches for and are thus likely constrained at similar levels to non-RPV models.
Of course, assuming a higher squark mass leads to shorter decay lengths, but in order to avoid a displaced vertex even for the squark LSP one has to go to high and hence quite unnatural squark masses.  It is essentially impossible to avoid a displaced vertex for a neutralino LSP.  This is all assuming baryogenesis happens around or above the weak scale.
Lower baryogenesis temperatures allow for shorter decay lengths and lighter squarks; however, even for 
baryogenesis occurring at 50 GeV, one is still forced to assume a squark mass above roughly 400 GeV for prompt decays; the minimum squark mass for prompt decays with 100 GeV baryogenesis is roughly 800 GeV. 

\begin{figure}
  \centering
  \includegraphics[scale=0.65]{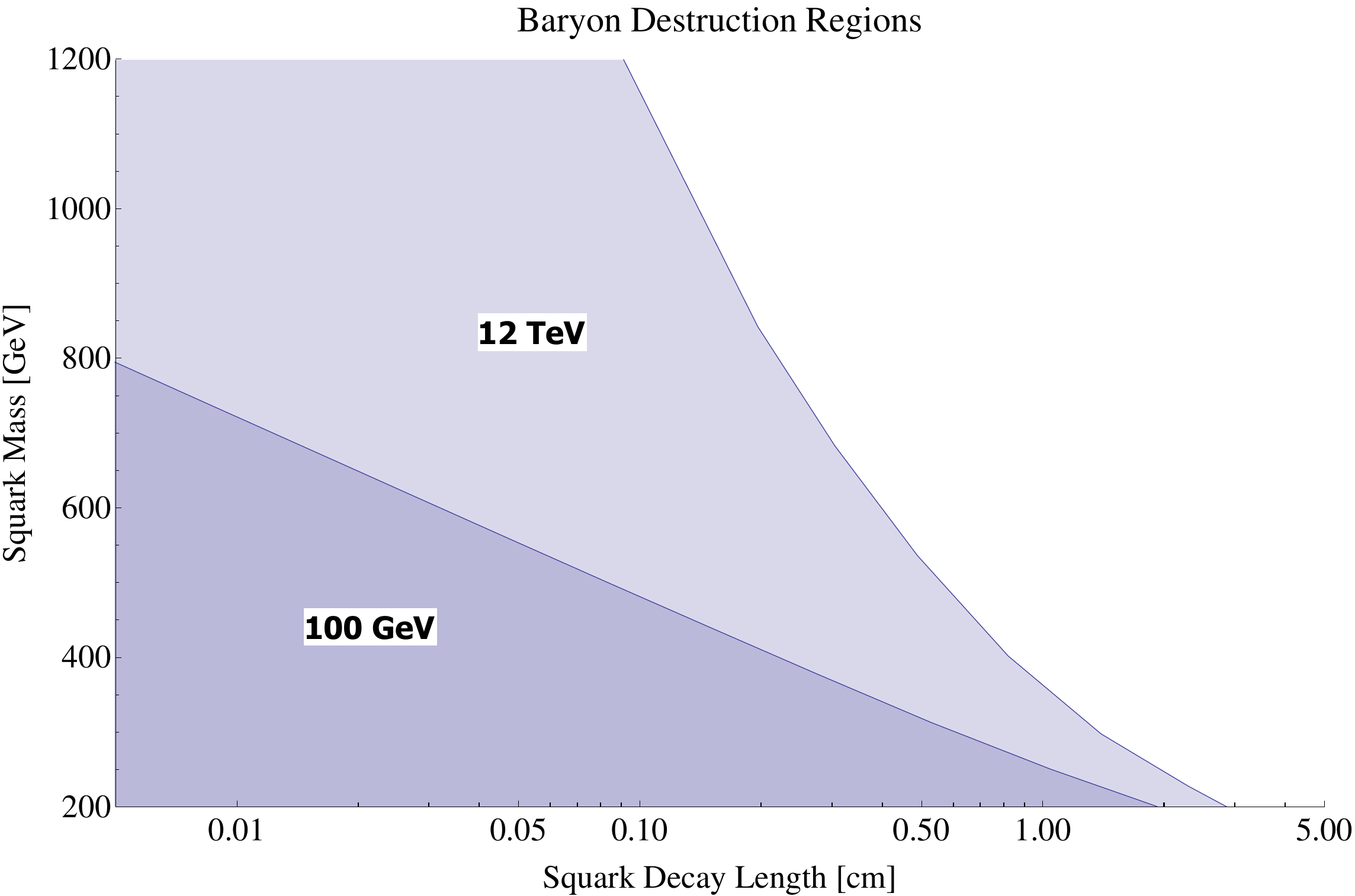}
  \caption{Shaded exclusion regions in the squark mass vs. squark decay length plane for a squark LSP with the baryon asymmetry generated at temperatures of 100 GeV and 12 TeV. For high scale models of baryogenesis, this shows that one expects squark decay lengths on the order of a centimeter or longer. For baryogenesis done at 100 GeV, squarks need to be heavier than 800 GeV to have prompt decays.}
  \label{DeathRegionMassVsDecayLength}
\end{figure}

To more clearly see the implications of these results for current LHC searches for \BPV processes, it is useful to view the exclusion regions on the squark mass versus LSP decay length plane, the latter parameter being much more experimentally relevant than the value of the \BPV coupling. Figure 
\ref{DeathRegionMassVsDecayLength} shows this for a squark LSP with an order-one baryon asymmetry generated at 12 TeV and 100 GeV starting temperatures. For the high scale exclusion region, one sees that the squark decay length should exceed a tenth of a centimeter. For an asymmetry generated at 100 GeV, the squark mass must 
not be less than 800 GeV, which corresponds to a 60 micron decay length, i.e.\ already a displaced vertex (based on the ATLAS displacement minimum). Since electroweak symmetry breaking occurs at roughly 100 GeV, this implies that any baryogenesis mechanism that happens at or before the electroweak phase transition will have displaced vertices in all SUSY events at the LHC, unless the squarks are very heavy ($\gtrsim 800$ GeV).
If a neutralino is the LSP then the decay lengths will be significantly longer, as in Figure \ref{DeathRegionMassVsLambdaNeutralino}.

\begin{figure}
  \centering
  \includegraphics[scale=0.55]{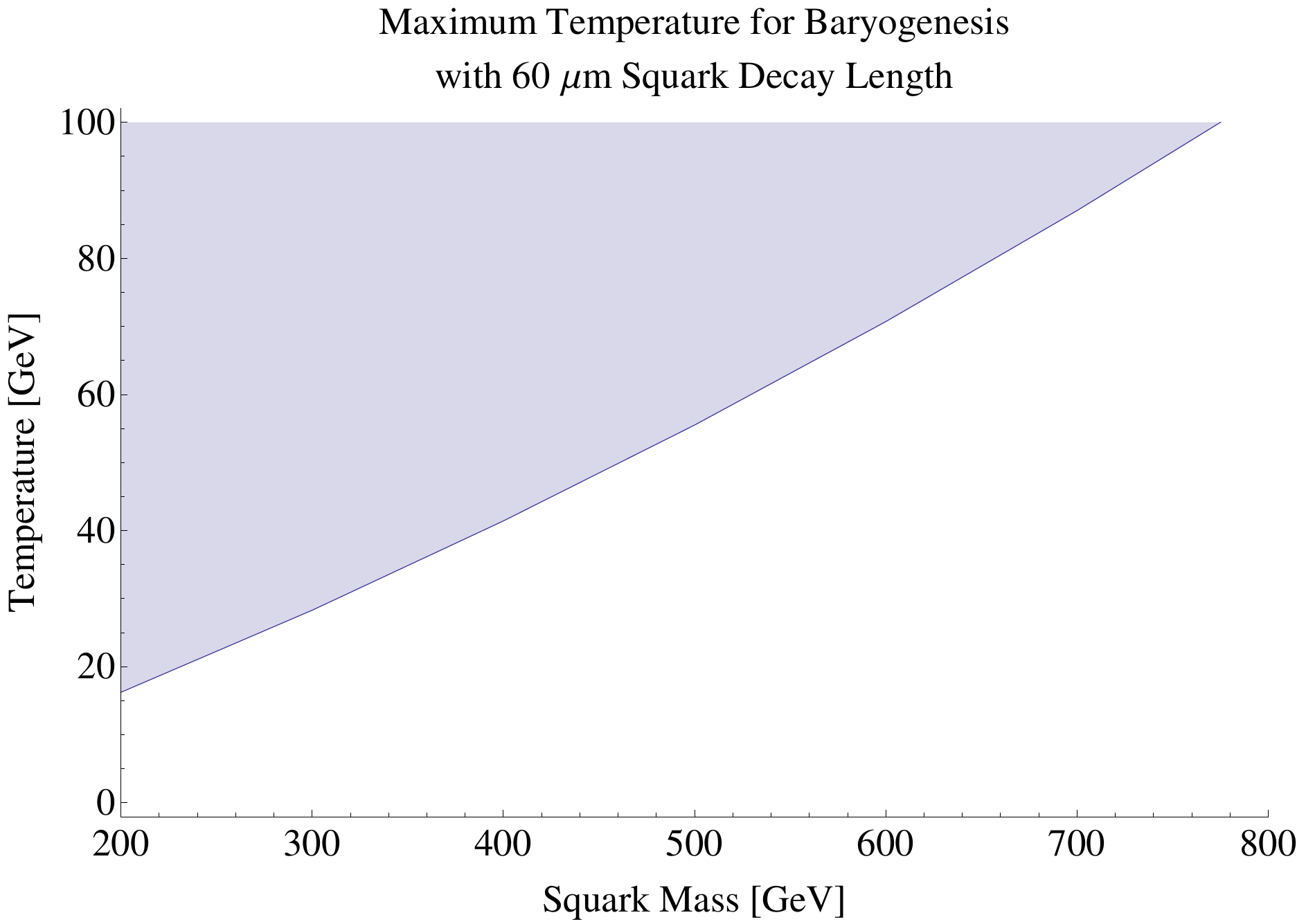}
  \caption{Assuming a squark LSP, the temperature below which a baryon asymmetry must be generated to avoid total destruction, given a 60 micron squark decay length (i.e.\ an ATLAS displaced vertex). The shaded region is excluded.}
  \label{MaxTemp60micronSquark}
\end{figure}

\begin{figure}
  \centering
  \includegraphics[scale=0.57]{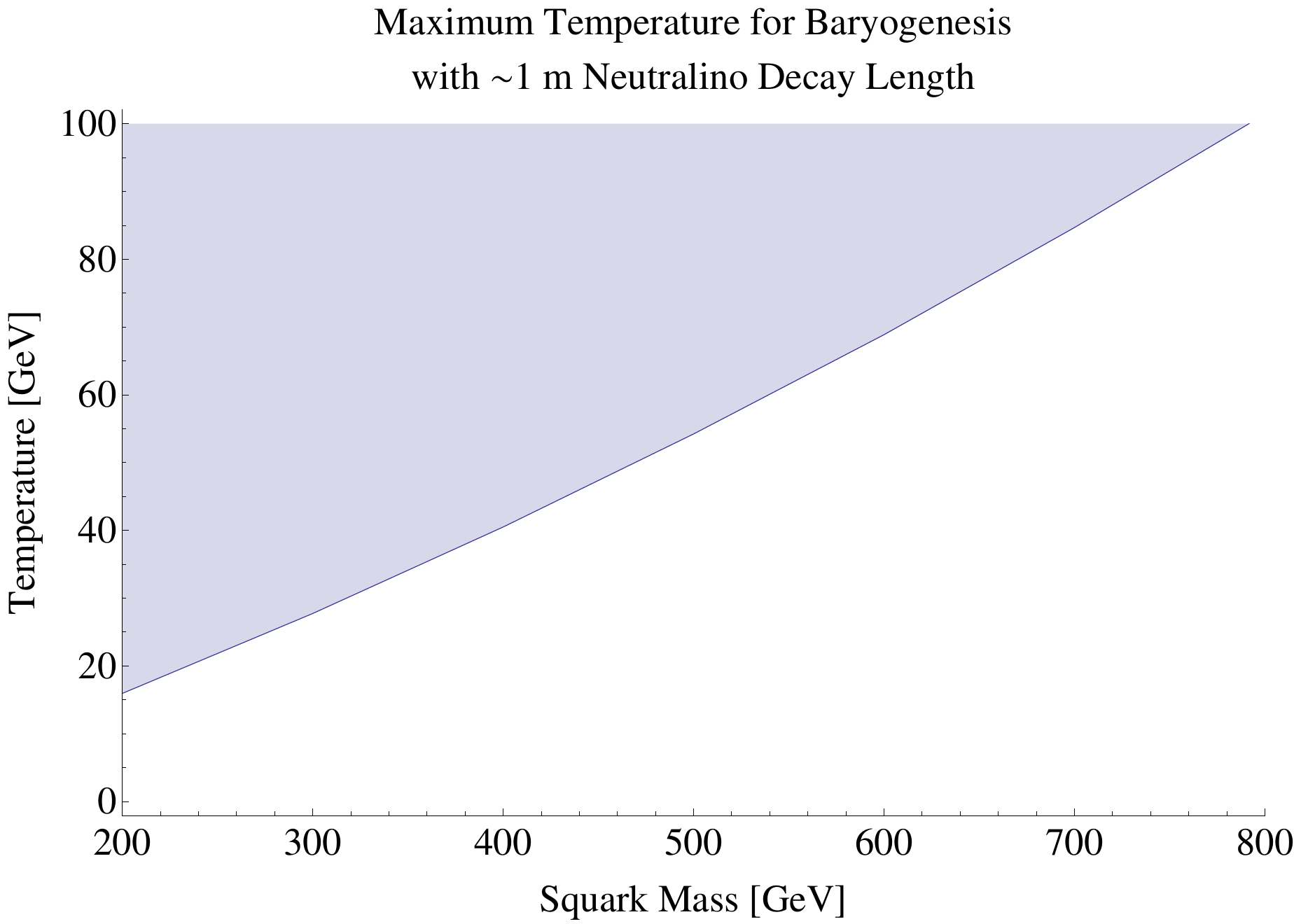}
  \caption{Assuming a neutralino LSP with a mass = 0.8 of the squark mass, the temperature below which a baryon asymmetry must be generated to avoid total destruction, given an approximately one meter neutralino decay length. The shaded region is excluded.}
  \label{MaxTemp1meterNeutralino}
\end{figure}

Figures \ref{MaxTemp60micronSquark} and \ref{MaxTemp1meterNeutralino} provide guidelines for the construction of models with \BPV and light squarks. The first assumes a squark LSP with a 60 micron decay length (the minimum criterion for a displaced vertex in the ATLAS detector) and plots the 
temperature below which a baryon asymmetry can be expected to survive as a function of the squark mass.  In other words to avoid having all SUSY events have displaced vertices, the \BPV model must have a baryogenesis mechanism below this temperature. The second figure shows similar information, this time for a neutralino LSP with an approximately 1 meter decay length.  Decay lengths much longer than this would appear in missing energy searches.  A model that over-generates baryons could avoid these constraints; however, this over-generation must be by many orders of magnitude, as we now explain. Our initial condition for the baryon asymmetry was about a million times larger than that of the actual universe. An 
asymmetry orders of magnitude larger than this would be needed to significantly affect the exclusion regions, which are not highly sensitive to this value. In the squark mass versus \(\log{|\lambda''|}\) plane (see e.g.\ Figure \ref{DeathRegionMassVsLambdaSquark}), low 
temperature exclusions appear as straight lines. The mass intercepts of these lines depend only logarithmically on the initial baryon asymmetry (the slopes are unaffected). To give a numerical measure of this effect, at a starting temperature of 100 GeV, changing the initial baryon asymmetry by a 
factor of 10 changes the mass-intercept of the exclusion region by roughly 12.5 GeV. The effect is linear in initial temperature, so for example, at a temperature of 50 GeV the change is only about 5.75 GeV. This weak dependence on initial baryon asymmetry ensures that our results are generic for essentially any model of baryogenesis.

\begin{figure}
  \centering
  \includegraphics[scale=0.5]{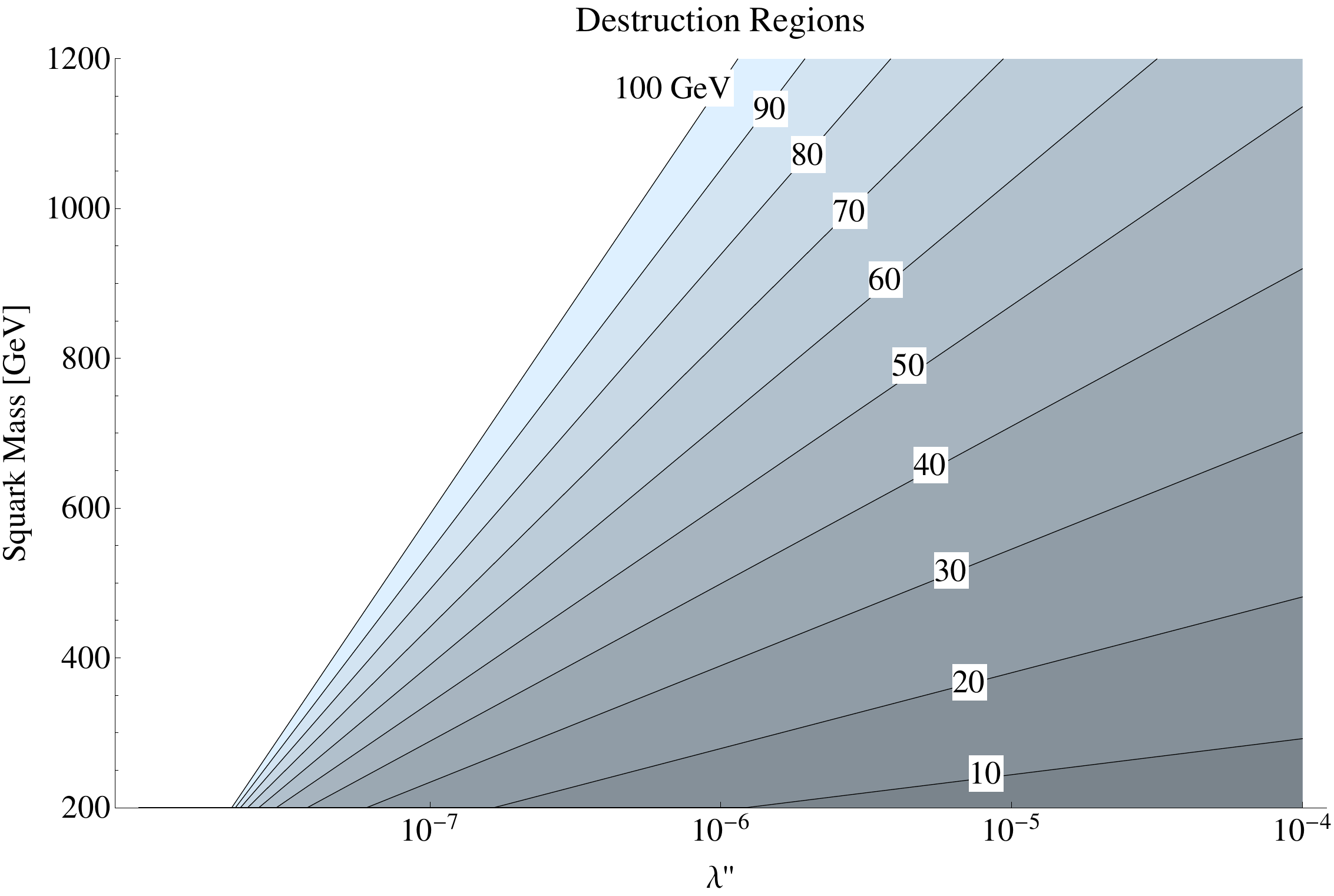}
  \caption{Exclusion regions for different starting temperatures below 100 GeV. Baryogenesis models in which baryon number is generated at a temperature T between 10 GeV and 100 GeV must assume squark mass and \(\lambda''\) values outside of the region shaded for that temperature.}
  \label{TempContours}
\end{figure}

Figure \ref{TempContours} plots the exclusion region for a variety of starting temperatures, providing more detailed guidance for model building than figures \ref{MaxTemp60micronSquark} and \ref{MaxTemp1meterNeutralino}. For example, suppose one wished for 400 GeV squarks and prompt decays. 
Effectively, all decays to the left of \(\log|\lambda''| = -6\) will have a significant fraction of displaced events, so 400 GeV squarks and prompt decays conservatively imply that baryogenesis should occur below about 30 GeV. The slopes of the lines in this plot are roughly predicted by Section 
\ref{GreatExpectations}, with the prediction for the lowest starting temperatures being the most accurate.

\section{Conclusions}
LHC searches have put strong constraints on superpartner masses, creating a tension with naturalness of the weak scale.  These constraints can be reduced, and the tension at least partially relieved by allowing R-parity violating operators, and in particular the $UDD$ operator.
However, this operator violates baryon number and hence in the early universe will lead to washout of the baryon asymmetry.  This process is very efficient in the early universe, and thus this coupling must be small to avoid this effect.  This naturally leads to long decay lengths for the LSP.  We have calculated the limits that the existence of the baryon asymmetry puts on the coupling strength.  Our main results are in Figures \ref{DeathRegionMassVsLambdaSquark}, \ref{DeathRegionMassVsLambdaNeutralino}, and \ref{DeathRegionMassVsDecayLength}.  We find that whenever the baryon asymmetry is generated around the weak scale or above, SUSY events at the LHC will likely all end in displaced vertices. Thus displaced vertices are a generic expectation of having the baryon-number violating RPV coupling in the model. If a particle other than the squark is the LSP, the decay will be very displaced, often on the scale of meters, and may in fact appear at the LHC as missing energy (or for example a long-lived CHAMP). In fact, for neutralino LSPs these decay lengths are long enough that in many cases they may already be ruled out by searches for missing energy at the LHC (see Figures \ref{DeathRegionMassVsLambdaNeutralino} and \ref{DeathRegionMassVsLambdaNeutralino2}), unless baryogenesis was performed at temperatures well below the weak scale.

Having every SUSY decay chain end in a displaced vertex can significantly affect the efficiency of searches for SUSY.
It is therefore important to design RPV searches at the LHC that are sensitive to displaced vertices.  The displaced vertices generally produce two or three jets, depending on whether or not the squark is the LSP.  They may or may not have a track leading to them, depending on whether the LSP is charged or neutral.  The displacement may range from $\sim 60 \mu$m to several meters and beyond.  Of course, depending on where in the LHC's detectors the decay occurs, the signals at the LHC are quite different.  Thus it is important to design searches that can see all the different decay lengths. Note that one may have a gravitino as the LSP, and our results apply to this case as well, so long as it is weakly enough coupled to the rest of the superpartners that the dominant decay of the NLSP is via the RPV coupling and not to the gravitino.

We are aware of only one relevant experimental search, done by the CMS collaboration \cite{CMSsearch}. This search sought displaced dijets from the decays of long-lived neutral particles, and placed limits on the 
production cross section of a process of the form \(H \rightarrow 2\chi \rightarrow 2q2\bar{q}\), where \(H\) is a non-Standard Model Higgs and the \(\chi\) is a netural boson with a displaced decay. These results have some 
applicability to both scenarios that we consider (squark and neutralino LSPs). Unfortunately, in both cases there are important factors that obfuscate the relative efficiency of this search. For this reason, we are unable 
to draw the experimental exclusions on our plots along with the theoretical exclusions. In the case of a neturalino LSP, the change in decay topology from two jets to three is the primary obstacle. This motivates searches 
specifically for three jet events. For a squark LSP, the main issue is the nature of the initial track. In general, the squark will form an R-hadron which may be charged or neutral, and may change charges as it 
travels through the detector. If the initial squark leaves a track, there will be little sensitivity to decay lengths over 50 cm, and in any event little sensitivity to decay lengths less than 500 microns \cite{Andrzej}. The efficiency for decay lengths between these two cutoffs is unknown.

The CMS exclusions are strong enough that, if the efficiency losses are not too large, light squark or neutralino LSPs may be ruled out at decay lengths greater than a millimeter or so. There should be a window at 
small decay lengths (60 to 500 microns) for two reasons. First, the lower limit on displaced vertices in \cite{CMSsearch} was 500 microns. Second, in our scenarios the decaying particles will be relatively less boosted than 
in the model used by \cite{CMSsearch}, meaning that at short decay lengths, more decays will fall below the detection cutoff, weakening limits. There may also be a window for squark LSPs at decay lengths greater than 50 cm, 
and in general the effect of the potentially charged nature of a squark track is hard to judge.
While it is not clear how strongly this particular search constrains our scenario, the strong limits placed by this search do demonstrate that a search that is designed for \BPV scenarios with displaced vertices would cover a large amount of the theoretically best-motivated parameter space.  The theoretical constraints that we have considered push \BPV scenarios into the long displaced vertex regime, which should be possible to cover well with LHC searches.

The only way to avoid our conclusion that the RPV coupling $UDD$ in combination with light squarks will lead to a displaced vertex is to generate the baryon asymmetry of the universe at a very low scale, below the 
electroweak scale. Electroweak baryogenesis or leptogenesis (that relies on sphalerons for example) will not work since enough of the squarks will remain below the weak scale that they will wash-out the baryon asymmetry. 
In fact, it has previously been shown \cite{Curtin:2012aa,Cohen:2012zza} that electroweak baryogenesis is ruled out in the MSSM due to the observed Higgs properties; our work shows that any new physics designed to 
resolve these issues must operate below the weak scale in the presence of prompt $UDD$ decays. In other words, if \BPV SUSY is discovered at the LHC without collider-scale displaced vertices, we would know that 
baryogenesis occurred at temperatures below the weak scale.

\section*{Acknowledgments}
We would like to thank Lawrence Hall, Jeremy Mardon, David Pinner, Wells Wulsin, Yue Zhao, and Andrzej Zuranski.
Kurt Barry acknowledges the support of a Stanford Graduate Fellowship during this work.
SR was supported by ERC grant BSMOXFORD no. 228169.  PWG acknowledges the support of the Hellman Faculty Scholars program and the Terman Fellowship.

\appendix

\section{Calculational Details}
\label{Sec:Appendix}
\subsection{Exact Expressions for Matrix Elements}
\begin{figure}{h}
  \centering
  \includegraphics[scale=0.65]{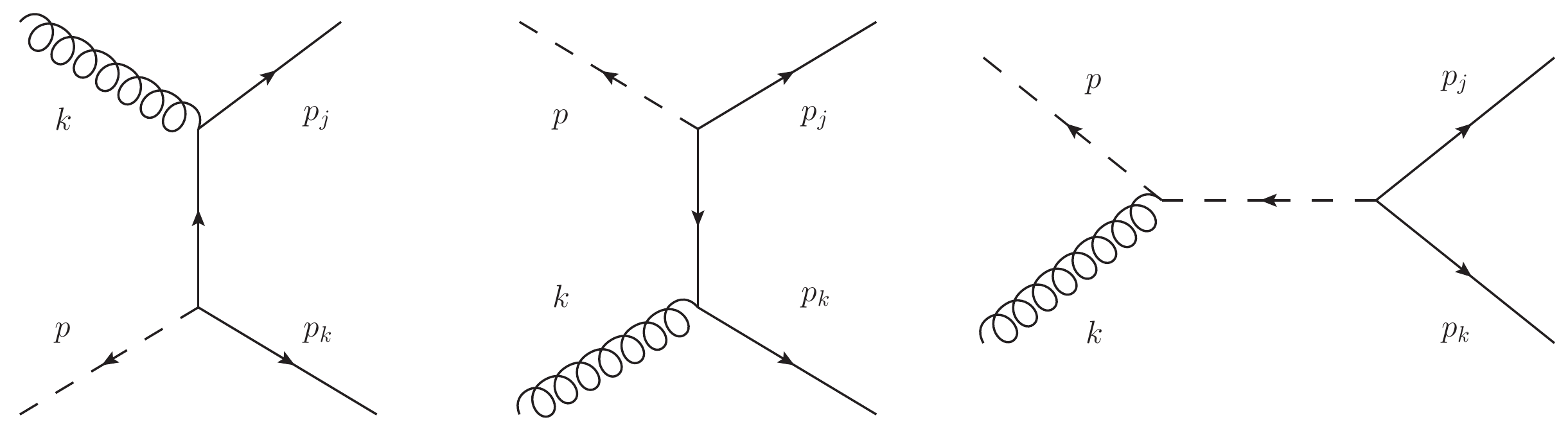}
  \caption{\textbf{\footnotesize{All diagrams contributing to Stimulated Decay of squarks, with momentum labels. The flow of time is left to right.}}}
  \label{StimDecAll}
\end{figure}

\begin{figure}{h}
  \centering
  \includegraphics[scale=0.65]{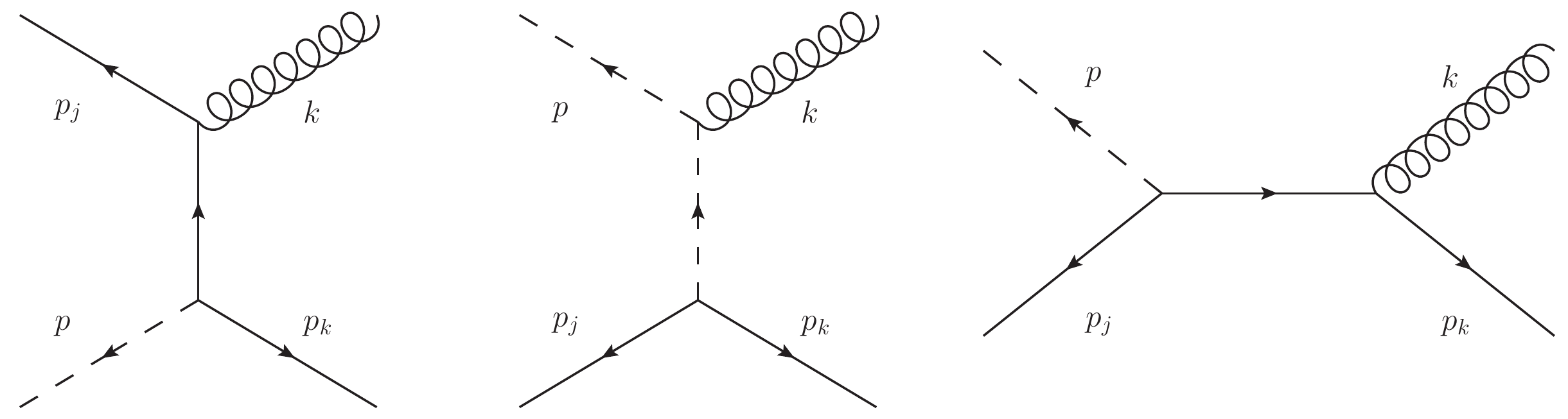}
  \caption{\textbf{\footnotesize{All diagrams contributing to Absorption of squarks, with momentum labels. The flow of time is left to right.}}}
  \label{AbsorbAll}
\end{figure}

The matrix elements quoted are summed over final state degrees of freedom and averaged over the initial state degrees of freedom. Equation \ref{MEStimDec} gives the squared matrix element for the 
Stimulated Decay process; the contributing diagrams, with their momentum labels, can be seen in figure \ref{StimDecAll}. Likewise, equation \ref{MEAbsorb} gives the squared matrix element for the Absorb 
process; the contributing diagrams are in figure \ref{AbsorbAll}. Since the difference between QCD and QED processes is only in the coefficients of the various terms, Table \ref{tab: coefficients}  summarizes the factors; the table applies to both matrix elements.  These matrix elements were calculated manually as the currently available symbolic matrix element calculating programs lack a built-in \(UDD\)-type vertex. The numerical results derived from these formulae were found to be in 
excellent agreement with order-of-magnitude estimates and displayed the correct approximate parametric behavior (cross sections go as \(1/T^2\) at high temperatures, and those with final state squarks are suppressed by a 
factor of \(\exp{-m/T}\) at low temperatures).

Stimulated Decay squared, summed matrix element:
\begin{eqnarray}
\nonumber
\sum |\mathcal{M}|^2 & = &  A e^2 |\lambda''|^2 
\frac{\left[2(p \cdot p_k)(p \cdot p_j) - m^2(p_j \cdot p_k) + m_k^2(p_j \cdot p_k - 2 p \cdot p_j)\right]}{\left(m^2 - 2 p \cdot p_k + m_k^2 - m_j^2\right)^2} \\ \nonumber
&& +  B e^2 |\lambda''|^2 
\frac{\left[2(p \cdot p_j)(p \cdot p_k) - m^2(p_k \cdot p_j) + m_j^2(p_k \cdot p_j - 2 p \cdot p_k)\right]}{\left(m^2 - 2 p \cdot p_j + m_j^2 - m_k^2\right)^2} \\ \nonumber
&& +  C e^2 |\lambda''|^2 
\frac{(p_j \cdot p_k)(m^2 + p \cdot k)}{4(p \cdot k)^2  + m^2\Gamma^2} \\ \nonumber
&& +  D e^2 |\lambda''|^2 
\frac{(p \cdot p_k - m_k^2)(p \cdot p_j - m_j^2)}{(m^2 - 2 p \cdot p_k + m_k^2 - m_j^2)(m^2 - 2 p \cdot p_j + m_j^2 - m_k^2)} \\ \nonumber
&& +  E e^2 |\lambda''|^2 
\frac{(p \cdot k)\left[ -m_k^2(m_j^2 + p_j \cdot p_k + 2 p \cdot p_j) + m_j^2(p \cdot p_k) + (p_j \cdot p_k)(m^2 + 2 p \cdot p_k)\right]}
{(m^2 - 2 p \cdot p_k + m_k^2 - m_j^2)[4(p \cdot k)^2  + m^2\Gamma^2]}\\
&& +  F e^2 |\lambda''|^2 
\frac{(p \cdot k)\left[ -m_j^2(m_k^2 + p_k \cdot p_j + 2 p \cdot p_k) + m_k^2(p \cdot p_j) + (p_k \cdot p_j)(m^2 + 2 p \cdot p_j)\right]}
{(m^2 - 2 p \cdot p_j + m_j^2 - m_k^2)[4(p \cdot k)^2  + m^2\Gamma^2]}
\label{MEStimDec}
\end{eqnarray}

Absorb squared, summed matrix element:
\begin{eqnarray}
\nonumber
\sum |\mathcal{M}|^2 & = & \frac{16}{3} A e^2 |\lambda''|^2 
\frac{\left[2(p \cdot p_k)(p \cdot p_j) - m^2(p_j \cdot p_k) + m_k^2(p_j \cdot p_k - 2 p \cdot p_j)\right]}{\left(m^2 - 2 p \cdot p_k + m_k^2 - m_j^2\right)^2} \\ \nonumber
&& + \frac{16}{3} B e^2 |\lambda''|^2 
\frac{\left[2(p \cdot p_j)(p \cdot p_k) - m^2(p_k \cdot p_j) + m_j^2(p_k \cdot p_j + 2 p \cdot p_k)\right]}{\left(m^2 + 2 p \cdot p_j + m_j^2 - m_k^2\right)^2} \\ \nonumber
&& + \frac{16}{3} C e^2 |\lambda''|^2 
\frac{(p_j \cdot p_k)(m^2 - p \cdot k)}{4(p \cdot k)^2} \\ \nonumber
&& + \frac{16}{3} D e^2 |\lambda''|^2 
\frac{(p \cdot p_k - m_k^2)(p \cdot p_j + m_j^2)}{(m^2 - 2 p \cdot p_k + m_k^2 - m_j^2)(m^2 + 2 p \cdot p_j + m_j^2 - m_k^2)} \\ \nonumber
&& + \frac{16}{3} E e^2 |\lambda''|^2 
\frac{-(p \cdot k)\left[ -m_k^2(-m_j^2 + p_j \cdot p_k + 2 p \cdot p_j) - m_j^2(p \cdot p_k) + (p_j \cdot p_k)(m^2 + 2 p \cdot p_k)\right]}
{(m^2 - 2 p \cdot p_k + m_k^2 - m_j^2)[4(p \cdot k)^2]}\\
&& + \frac{16}{3} F e^2 |\lambda''|^2 
\frac{-(p \cdot k)\left[ -m_j^2(-m_k^2 + p_k \cdot p_j - 2 p \cdot p_k) + m_k^2(p \cdot p_j) + (p_k \cdot p_j)(m^2 - 2 p \cdot p_j)\right]}
{(m^2 + 2 p \cdot p_j + m_j^2 - m_k^2)[4(p \cdot k)^2]}
\label{MEAbsorb}
\end{eqnarray}

\begin{table}
\begin{tabular} {|c|c|c|c|c|c|c|c|c|}
\hline
  Gauge Boson & Squark Flavor & Quark j Flavor & A     & B      & C      & D      & E      & F      \\ \hline
  gluon       & n/a           & n/a            & 8/3   & 8/3    & -16/3  & 16/3   & -8/3   & -8/3   \\ \hline
  photon      & up            & down           & 4/27  & 4/27   & -32/27 & -16/27 & -16/27 & -16/27 \\ \hline
  photon      & down          & up             & 16/27 & 4/27   & -8/27  & 32/27  & -16/27 & 8/27   \\ \hline
  photon      & down          & down           & 4/27  & 16/27  & -8/27  & 32/27  &   8/27 & -16/27 \\
  \hline
\end{tabular}
\caption{Coefficients for the processes in equations \eqref{MEStimDec} and \eqref{MEAbsorb}}
\label{tab: coefficients}
\end{table}

\subsection{Other Approximations}
\begin{itemize}

\item
  Maxwell-Boltzmann statistics were used in thermal averages instead of proper quantum distribution functions. This choice was crucial for rendering the integrals tractable.

\item
  In a thermal environment, one should properly use thermally corrected Feynman rules and sum over a broader range of diagrams to avoid IR divergences. Here, IR divergences were instead
  regulated with a simple cutoff (only two of five processes were affected by this decision). The resulting cross sections have the correct order of magnitude and their thermal 
  averages display the correct scaling with temperature, so this choice is believed to be reasonable. A detailed discussion of thermal corrections may be found in \cite{Kamionkowski}.

\item
  Exchange terms were assumed to enforce perfect equality of different species--in reality there is a small difference, but is roughly as small as the baryon destruction rates compared to the exchange 
  rates.

\item
  The effective degrees of freedom of the universe (\(g_*\)) was taken to have a constant value of 100--an approximation with error not worse than 10 percent over the range of temperatures considered.

\item
  Processes involving only virtual squarks were neglected as these higher dimensional operators should be insignificant.

\item
  Difference in decay rates between up and down squarks were neglected in resonance terms (numerical experiments showed this to be insignificant).

\item
  Fixed quark masses were used to regulate co-linear divergences (this was also shown to be insignificant with numerical experiments).

\item
  The effective entropic degrees of freedom, \(g_{*S}\), were taken to be approximately equal to the effectively massless degrees of freedom, \(g_*\) (a good approximation as long as all particle species 
  have roughly the same temperature).

\item
  The Higgs contribution to baryon number destruction from processes involving initial and final state Higgs particles was ignored to avoid having to track Higgs abundance as well. The smallness of the Yukawa couplings makes this negligible.

\item
  Weak processes have been neglected. This is unlikely to have a large impact, since QCD processes are dominant due to the ratio of the couplings and the larger number of gluons than W and Z bosons.
  
\item
  Electromagnetic processes were calculated, but were found to be negligible (though they were still included in the numerical calculation).
  
\item
  The matrix elements were only calculated to tree level.

  
\end{itemize}

\end{document}